\documentclass[apj]{emulateapj}

\usepackage{apjfonts}
\usepackage{epsfig}
\usepackage{graphicx}
\usepackage{url}
\usepackage{natbib}
\usepackage{float}
\usepackage{amsmath}
\usepackage[bf, sf, FIGTOPCAP, nooneline, tight]{subfigure}
\usepackage{graphicx}
\usepackage{dcolumn}
\usepackage{bm}
\usepackage[usenames,dvipsnames]{color}
\usepackage[citebordercolor={0 1 0}]{hyperref}
\usepackage{amssymb}

\newcommand{\be}{\begin{equation}}
\newcommand{\ee}{\end{equation}}
\newcommand{\bq}{\begin{eqnarray}}
\newcommand{\eq}{\end{eqnarray}}

\def\({\left(}
\def\){\right)}

\begin{document}

\title{Cosmological constraints from the redshift dependence of the
Alcock-Paczynski effect: Fourier space analysis}

\author{Xiaolin Luo, Ziyong Wu, Xiao-Dong Li, Miao Li,}
\affiliation{School of Physics and Astronomy, Sun Yat-Sen University, Guangzhou 510297, P.R.China}
\email{Corresponding Authors:  lixiaod25@mail.sysu.edu.cn}

\author{Zhigang Li}
\affiliation{Nanyang Normal University, 1638 Wolong Rd, Wolong, Nanyang, China}



\author{Cristiano G. Sabiu}
\affiliation{Department of Astronomy, Yonsei University, 50 Yonsei-ro, Seodaemun-gu, Seoul, 03722, Korea}



%


\begin{abstract}
The tomographic Alcock-Paczynski (AP) method 
utilizes the redshift evolution of the AP distortion
to place constraints on cosmological parameters.
It has proved to be a robust method that can separate the AP
signature from the redshift space distortion (RSD) effect,
and deliver powerful cosmological constraints using the $\lesssim 40h^{-1}\ \rm Mpc$ clustering region.
In previous works, the tomographic AP method was performed via the anisotropic 2-point correlation function statistic. 
In this work we consider the feasibility of conducting the analysis in the Fourier domain and examine the pros and cons of this approach. 
We use the integrated galaxy power spectrum (PS) as a function of direction, $\hat P_{\Delta k}(\mu)$,
to quantify the magnitude of anisotropy in the large-scale structure clustering,
and use its redshift variation to do the AP test.
The method is tested on the large, high resolution Big-MultiDark Planck (BigMD) simulation
at redshifts $z=0-1$, using the underlying true cosmology $\Omega_m=0.3071,\ w=-1$.
Testing the redshift evolution of $\hat P_{\Delta k}(\mu)$ in the true cosmology and cosmologies
deviating from the truth with $\delta \Omega_m=0.1,\ \delta w=0.3$,
we find that the redshift evolution of the AP distortion overwhelms the effects created by the RSD by a factor of $\sim1.7-3.6$.
We test the method in the range of $k\in(0.2,1.8)\ h\ \rm Mpc^{-1}$,
and find that it works well throughout the entire regime.
We tune the halo mass within the range $2\times 10^{13}$ to  $10^{14}\ M_{\odot}$,
and find that the change of halo bias results in $\lesssim 5 \%$ change in  $\hat P_{\Delta k}(\mu)$,
which is less significant compared with the cosmological effect.
Our work shows that it is feasible to conduct the tomographic AP analysis in the Fourier space.
\end{abstract}

\maketitle







\section{Introduction}\label{intro}

The large-scale structure (LSS) of the Universe contains
enormous information about the expansion and structure growth histories of our Universe.
In the past two decades, large-scale surveys of galaxies
has greatly enriched our understanding about the Universe
\citep{york2000sloan,2df:Colless:2003wz,Eisenstein:2005su,Percival:2007yw,blake2011wigglez,blake2011wigglezb,beutler20116df,
anderson2012clustering,alam2017clustering},
while the future surveys will enable us to measure
the $z\lesssim1.5$ Universe in unprecedented precision,
shedding light on the dark energy problem \citep{riess1998observational,perlmutter1999measurements,
weinberg1989cosmological,miao2011dark,weinberg2013observational}.



The well-known Alcock-Paczynski (AP) test \citep{Alcock:1979mp} is a geometric method for
probing the cosmic expansion history using the LSS. 
Under a certain cosmological model, the radial and tangential sizes of distant objects or structures take the forms of
$\Delta r_{\parallel} = \frac{c}{H(z)}\Delta z$ and $\Delta r_{\bot}=(1+z)D_A(z)\Delta \theta$,
where $\Delta z$, $\Delta \theta$ are their redshift span and angular size,
while $D_A$, $H$ being the angular diameter distance and the Hubble parameter, respectively.

Assuming incorrect models for computing Da and H results in miss-estimated values of $\Delta r_{\parallel}$ and $\Delta r_{\bot}$,
which manifest themselves as geometric distortions along the line-of-sight (LOS) and perpendicular directions. 
This distortion, known as the AP distortion, can be measured and quantified
via statistical analysis of the large-scale galaxy distribution,
and thus is widely used in LSS survey analyses to constrain the cosmological parameters
\citep{ryden1995measuring,Ballinger1996,matsubara1996cosmological,outram20042df,blake2011wigglez,lavaux2012precision,alam2017clustering,
Qingqing2016,KR2018}.

\begin{figure*}[htbp]
      \centering
      \includegraphics[width=9.4cm]{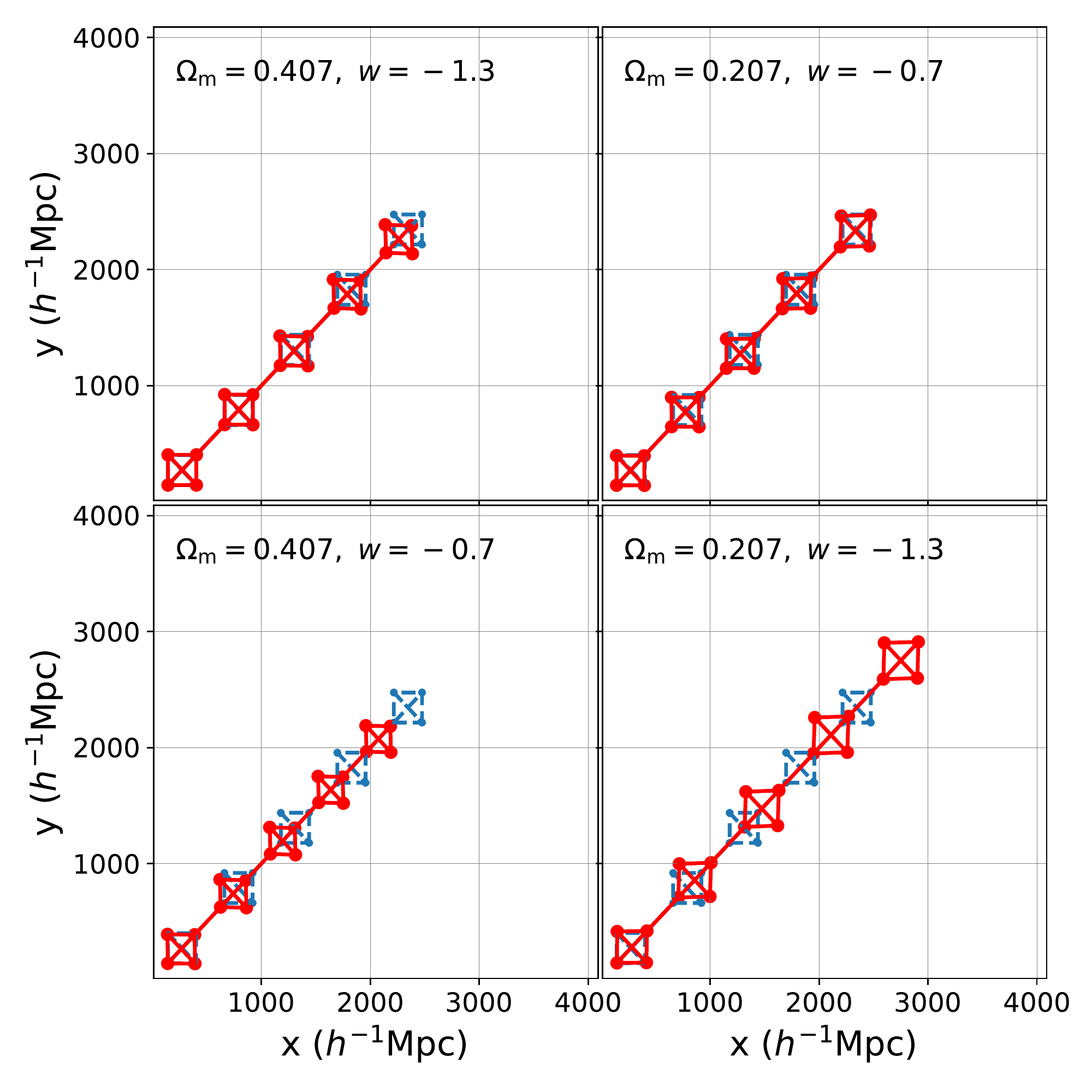}
      \includegraphics[width=8.5cm]{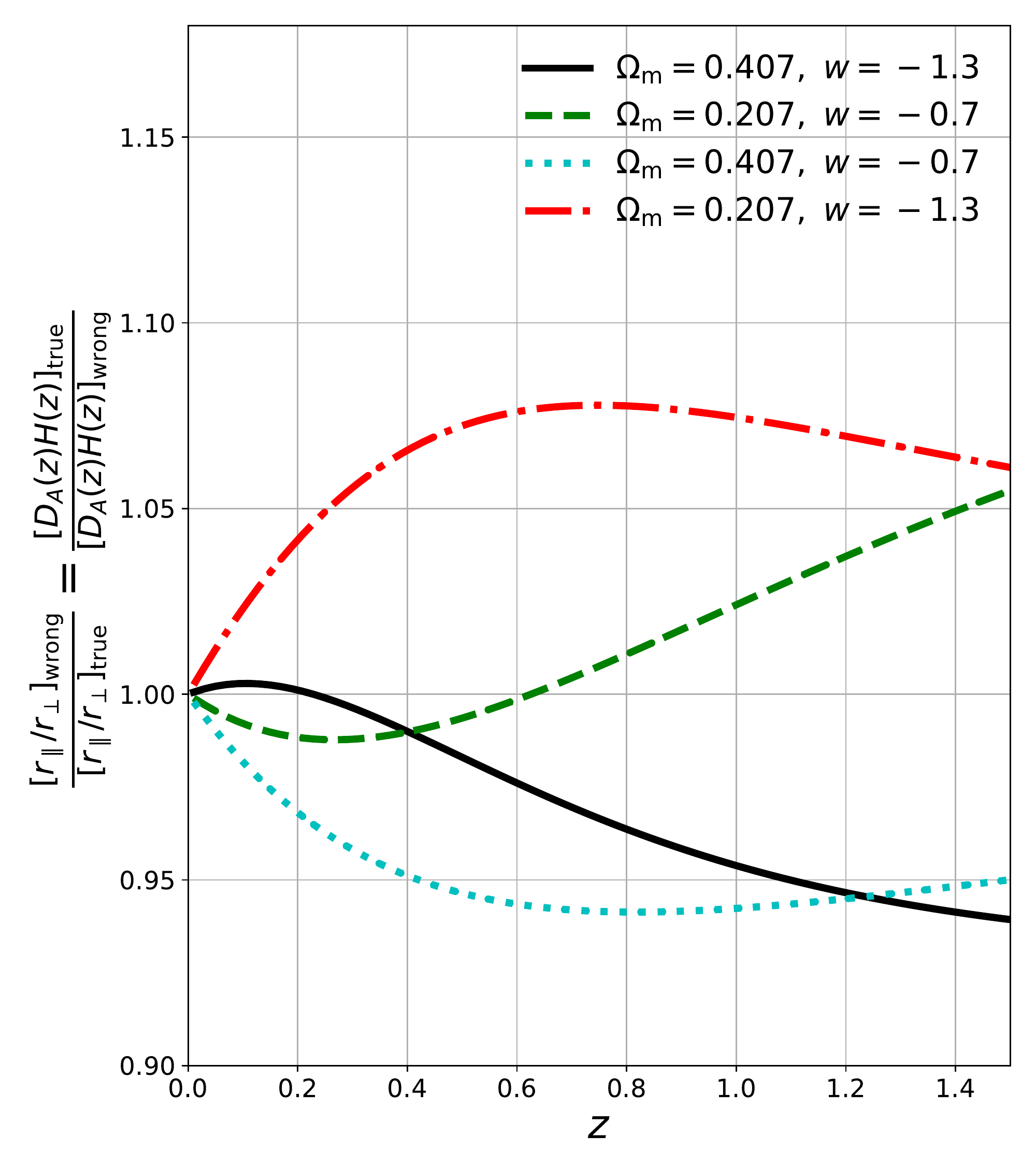}
      \label{fig_xy}
      \caption{Examples of shapes distorted due to the AP effect,
      in the case of adopting four incorrect sets of cosmologies
      in a fiducial cosmology of $\Omega_m=0.307$, $w=-1$.
      In the left panel, we plot the underlying true shapes of the 5 perfect squares with blue dashed lines,
      while their distorted shapes in the wrong cosmologies are showed by red solid lines.
      The observer was placed at the origin of the 2D coordinates.
      In the right panel we plot the magnitude of the shape distortions as a function of redshift
      (Equation \ref{eq_DAH}). }
      \label{fig:bias}
\end{figure*}




The ``tomographic AP method'' is a novel technique of applying the AP test to the LSS \citep{LI14,LI:2015jra,Park:2019mvn}, 
which has achieved tight constraints on the parameters governing the cosmic expansion.
The concept of the method is to utilize the {\it redshift evolution} of the LSS anisotropy,
which is sensitive to the AP effect, while being insensitive to the distortion produced by the redshift space distortions (RSD).
This makes it possible to differentiate the AP distortion from the large contamination of RSD.
\cite{LI:2015jra} proposed to quantify the anisotropic clustering via $\hat \xi_{\Delta s}(\mu)$,
which is defined as an integration of the 2D $\xi(s,\mu)$ over the clustering scale $s$.
\cite{Li:2016wbl} firstly applied the method to the
SDSS (Sloan Digital Sky Survey) BOSS (Baryon Oscillation Spectroscopic Survey) DR12 galaxies,
and achieved $\sim35\%$ improvements in the constraints
on the ratio of dark matter $\Omega_m$ and dark energy equation of state (EOS) $w$.
In later works, \cite{LI:2018nlh,Zhang2019} found the method greatly improves the
constraints on dynamical dark energy models,
while the analysis of \cite{LI19} showed that, in the case of using DESI-like data
\footnote{https://desi.lbl.gov/}, combining the method with the CMB+BAO datasets
can improves the dark energy figure-of-merit \citep{Wang:2008zh} by a factor of 10.

In this work, we extend the scope of the previous studies
and investigate how to conduct the tomographic AP test using the power spectrum (PS).
The PS is the Fourier transform of the two-point correlation function.
Compared to the correlation function in configuration space, the PS has advantages such as
milder coupling among the different $k$-modes, 
it is more closely related to theoretical models, and so on.
Thus it has been widely adopted as a standard tool in galaxy clustering analysis.
It is worthwhile developing a methodology to conduct the tomographic AP method in the Fourier space,
and the result will serve as a cross-check to the configuration space result.


The rest of this paper is organized as follows.
In Sec.\ref{AP}, we briefly introduce the physics of the AP test as well as our methodology.
In Sec.\ref{data}, we describe the N-body simulation and the halo samples used in this work.
We present the results in Sec.\ref{result}, and conclude in Sec.\ref{conclu}.

      \begin{figure*}[htbp]
      \centering

      \includegraphics[width=14cm,trim=130 80 10 80,clip]{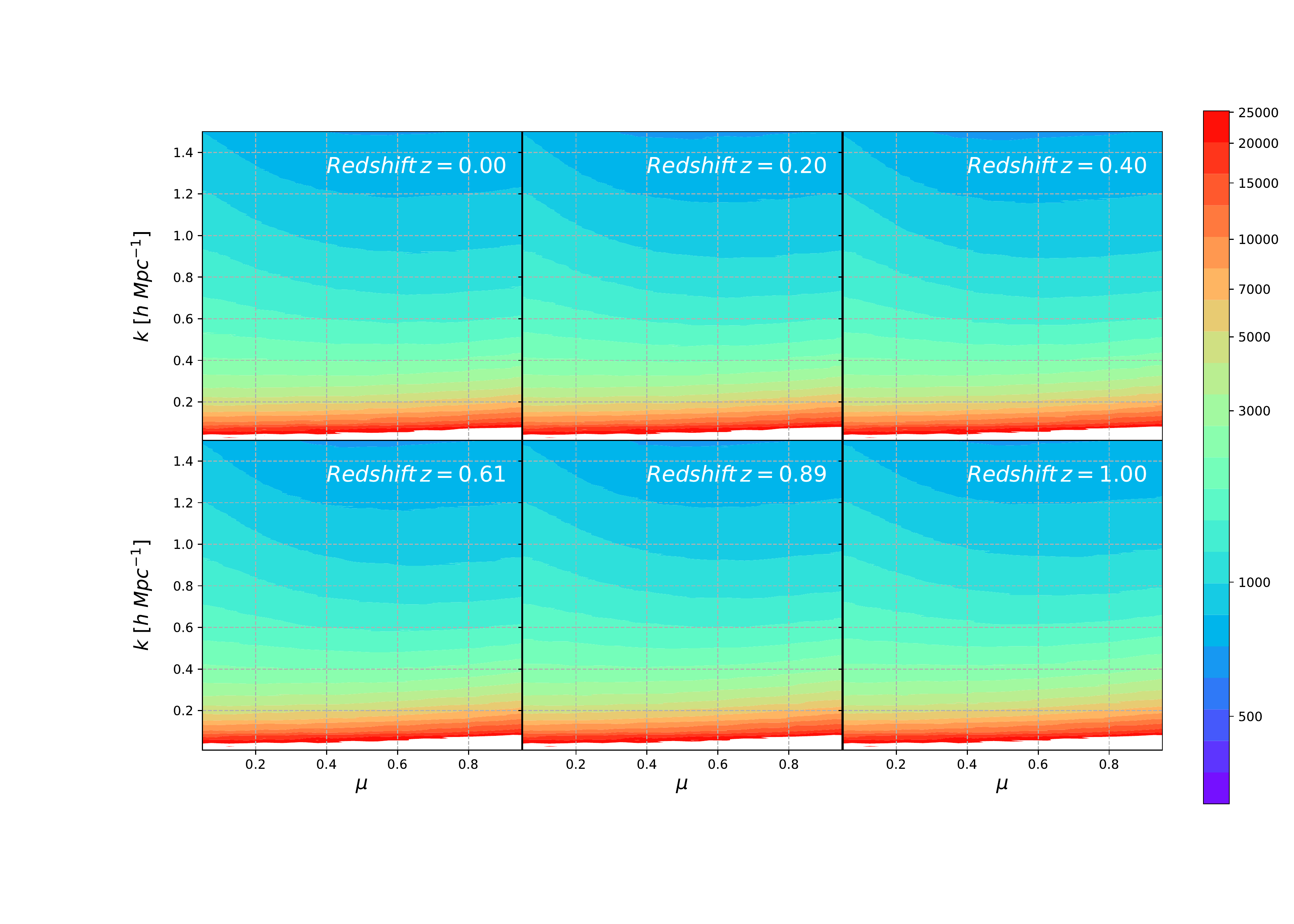}
      \caption{The 2D PS $P(k,\mu)$ of the BigMD halos measured
      at 6 redshifts distributed in $z\in [0,1]$,
      in the framework of the true cosmology (the simulation cosmology).
      The RSD effect produces anisotropies in the halo distribution,
      and manifests itself as the non-horizontal contour lines.
      The similarity among the 6 panels shows that in different redshifts
      the anisotropy created by the RSD maintains a similar pattern.
      }
      \label{fig:contour}
      \end{figure*}

\section{Methodology}\label{AP}


\subsection{The Alcock-Paczynski Test in a Nutshell}

The so called AP effect \citep{Alcock:1979mp} refers to
the apparent geometric distortions in the LSS
that arise when incorrect cosmological models
are assumed for transforming the observed galaxy redshifts into comoving distances.
For a distant object or structure in the Universe,
one can calculate its size along and across the LOS using the formulas of
\be
r_{\parallel}=\frac{c}{H(z)}\Delta z,\quad r_{\perp}=(1+z)D_A(z)\Delta\theta.
\ee
Here $\Delta z$ and $\Delta\theta$ are the observed redshift span and angular size measured via observations,
while $H$ and $D_A$ are the Hubble parameter and the angular diameter distance, respectively.
In a spatially flat Universe composed of a dark matter component with current ratio $\Omega_m$ and
a dark energy component with constant EOS $w$, 
they take the forms of
\be
H(z)=H_0\sqrt{\Omega_m a^{-3}+(1-\Omega_m)a^{-3(1+w)}},
\ee
\be
D_A(z)=\frac{1}{1+z}r(z)=\frac{1}{1+z}\int_0^z \frac{c dz'}{H(z')},
\ee
where $a=1/(1+z)$ is the cosmic scale factor, $H_0$ is the current Hubble parameter and $r(z)$ is the comoving distance.

Adopting a wrong set of $\Omega_m$ and $w$ results in miss-estimated  values of $r_{\parallel}$ and $r_{\perp}$.
Consequently, the constructed objects have distorted shapes (AP effect) and incorrect volume elements (volume effect),
whose magnitudes are
\be
\frac{{\rm Shape}_{\rm wrong}}{{\rm Shape}_{\rm true}}=
\frac{[r_{\parallel}/r_{\perp}]_{\rm wrong}}{[r_{\parallel}/r_{\perp}]_{\rm true}}=\frac{[D_A(z)H(z)]_{\rm true}}{[D_A(z)H(z)]_{\rm wrong}},
\label{eq_DAH}
\ee
\be
\frac{{\rm Volume}_{\rm wrong}}{{\rm Volume}_{\rm true}}=\frac{[(r_{\parallel}r_{\perp})^2]_{\rm wrong}}{[(r_{\parallel}r_{\perp})^2]_{\rm true}}=
\frac{[(D_A(z))^2/H(z)]_{\rm wrong}}{[(D_A(z))^2/H(z)]_{\rm true}}.
\ee
In galaxy surveys, measuring the galaxy clustering in the radial and transverse directions
leads to measurements of $D_A(z)$ and $H(z)$,
which enables us to place constraints on the cosmological parameters therein.

In Figure \ref{fig_xy} we illustrate the AP distortions
for four incorrect cosmologies.
The figure shows that, not only the shapes of the distributions are distorted,
but also a redshift dependence in the distortion appears in each cosmology.
Based on this fact, \cite{LI14,LI:2015jra} proposed the tomographic AP method,
which probes the AP distortion by measuring the redshift evolution of the
anisotropic galaxy clustering.

\subsection{Tomographic AP test using PS}

\cite{LI14} proposed a tomographic analysis of the small scale galaxy clustering to efficiently separate the AP effect from the RSD. 
They used the integrated correlation function at various LOS directions, $\hat \xi_{\Delta s}(\mu)$, 
to quantify the magnitude of anisotropy at different redshifts.
\cite{LI:2015jra} applied the method to the SDSS galaxy samples
and obtained tight cosmological constraints.
It is worthy to develop a similar method in Fourier space, as an alternative way to conduct the tomographic AP analysis.

Using the PS statistics has many advantages over using the correlation function in configure space, such as,
\begin{itemize}
  \item In the PS, clustering signals at different scales are uncorrelated, 
  while in configuration space there exists strong mode-coupling.
  By using the PS analysis we can more easily prohibit the clustering signal in the heavily non-linear region
   from entering a wide range of modes and causing arduous complexity.
  Also, being more clear about what physical scales are used in the analysis,
   we can have a better understanding about the method.
  \item Compared with the configuration space,
  in Fourier space it is more convenient to calculate the theoretical predictions of the statistical quantities.
  \item The computation of the PS can be much faster than the 2-point correlation function.
\end{itemize}

In this work, we adopt the open-source tool \texttt{Nbodykit} \citep{Hand:2017pqn} to calculate the 2D PS.
The redshift-space data  is assigned to a discrete mesh using the \texttt{to\_mesh} function with the default Cloud-In-Cell window,
and the values of $P(k,\mu)$ are then computed using the Fast Fourier Transform \citep{Bianchi:2015oia,Scoccimarro:2015bla}.
Here $k$ is the wavenumber, and $\mu$ represents the cosine of the angle between line of sight and wavenumber.

In conducting the AP test we first measure the 2D PS $P(k,\mu)$ of the BigMD sample. 
Following the concept of \cite{LI:2015jra}, we then integrate $P(k,\mu)$ over the wavenumber $k$,
to build up a statistical quantity that solely depends on the direction $\mu$,
\begin{equation}\label{intePS}
 P_{\Delta k}(\mu) \equiv \int_{k_{min}}^{k_{max}}P(k,\mu)dk.
\end{equation}
To reduce the effect of the clustering strength and galaxy bias, we further conduct a normalization,
which is expressed as
\begin{equation}
 \hat P_{\Delta k}(\mu) \equiv \frac{P_{\Delta k}(\mu)}{\int_0^1 P_{\Delta k}(\mu)d\mu}.
\end{equation}
\cite{LI:2015jra,Li:2016wbl} suggested using $s=6-40h^{-1}\rm Mpc$ for $\hat \xi_{\Delta s}(k,\mu)$.
This corresponds to a $k$ range of $\sim(0.15-1)\ h\ \rm Mpc^{-1}$.
In this work, we will test several choices of $(k_{\rm min}, k_{\rm max})$,
to gain some understanding about which range of $k$ is most optimal for this analysis. 


\section{Mock}\label{data}

For the testing of the new method we use the Big-MultiDark Planck (BigMD) simulation \citep{2016MNRAS.457.4340K},
whose size and resolution is close to the Horizon Run N-body simulations \citep{kim2009horizon,kim2015horizon} used in \cite{LI14,LI:2015jra,Li:2016wbl}.
The simulation was produced using $3\,840^3$ particles in a volume of $(2.5h^{-1}\rm Gpc)^3$,
assuming a $\Lambda$CDM cosmology with parameters
 $\Omega_m = 0.307115$, $\Omega_b = 0.048206$, $\sigma_8 = 0.8288$, $n_s = 0.9611$, and $H_0 = 67.77 {\rm km}\ s^{-1} {\rm Mpc}^{-1}$.
This yields a mass resolution of $2.4\times 10^{10} h^{-1}M_{\odot}$.
The initial condition was set by using the Zeldovich approximation
at redshift $z_{\rm init}=100$.
From the simulation particles, halo catalogues at 78 snapshots are
generated using the Rockstar algorithm \citep{2013ApJ}.
The large volume, high resolution resolution and wide redshift coverage of
the simulation makes it among
the best choices for testing our methodology.

The snapshots chosen for the Fourier space analysis of AP effect 
have redshifts of $0,0.2,0.4,0.61,0.89$ and $1.0$, respectively.
By applying different minimal mass cuts at different redshifts 
to maintain a constant number density $\bar n = 10^{-3}\ (h^{-1}\rm Mpc)^{-3} $
in all snapshots.
This number density is close to the galaxy number density of current and next generation
spectroscopic galaxy surveys.
To mimick the redshift-space distortions (RSD) caused by galaxy peculiar velocities,
we perturb the positions of halos along the $z$-direction,
using the following formula
\begin{equation}\label{eq:zvpeu}
\Delta z = (1+z) \frac{v_{{\rm LOS}}}{c},
\end{equation}
where $v_{\rm LOS}$ is the line-of-sight (LOS) component of the peculiar velocity of halos.

\begin{figure*}[htbp]
      \centering
      \includegraphics[width=14cm]{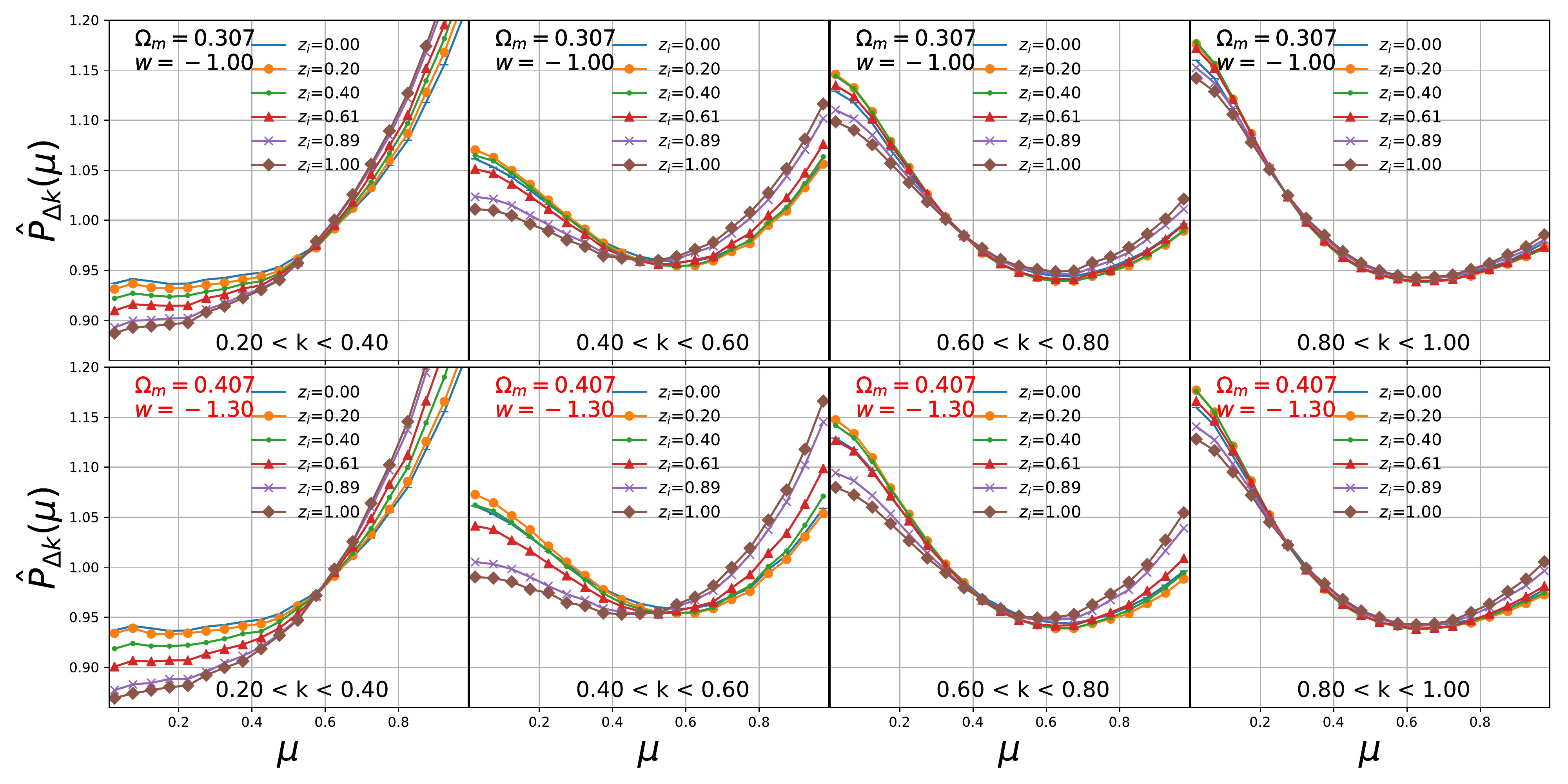}
      \caption{The integrated 2D PS $\hat P_{\Delta k}(\mu)$
      measured at 6 redshifts, in the underlying true cosmology (upper panels)
      and a wrong cosmology $(\Omega_m=0.407,w=-1.3)$ (text marked in red), respectively.
      In both cases we detect large anisotropy mainly produced by the RSD effect.
      In the wrong cosmology case,
      a larger redshift evolution of $\hat P_{\Delta k}(\mu)$
      produced by the AP distortion is detected.
      \label{fig:smallprime}}
      \end{figure*}

      \begin{figure*}[htbp]
      \centering
      \includegraphics[width=15cm]{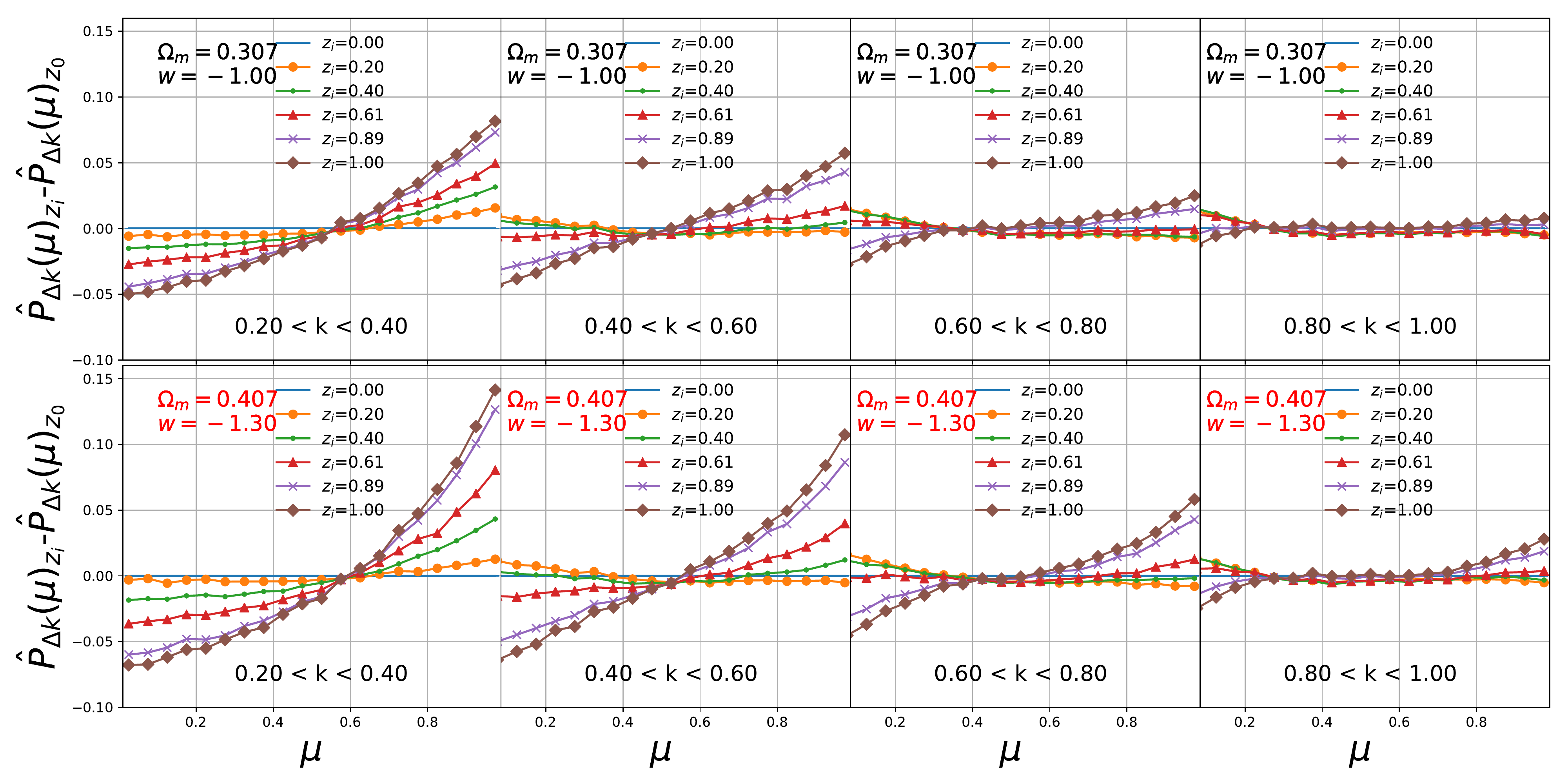}
      \caption{The redshift evolution of $\hat P_{\Delta k}(\mu)$
      between redshifts $z_i$ and $z_0$,
      where we chose $z_0=0$ and $z_i=0.2,0.4,0.61,0.89,1$.
      A larger redshift evolution is detected in the wrong cosmology.
      The integration range of $k$ is chosen as (0.2,0.4), (0.4,0.6), (0.6,0.8) and (0.8,1) $h\ \rm Mpc^{-1}$, respectively.
      }
      \label{fig:small}
      \end{figure*}

      \begin{figure*}[htbp]
      \centering

      \includegraphics[width=8cm]{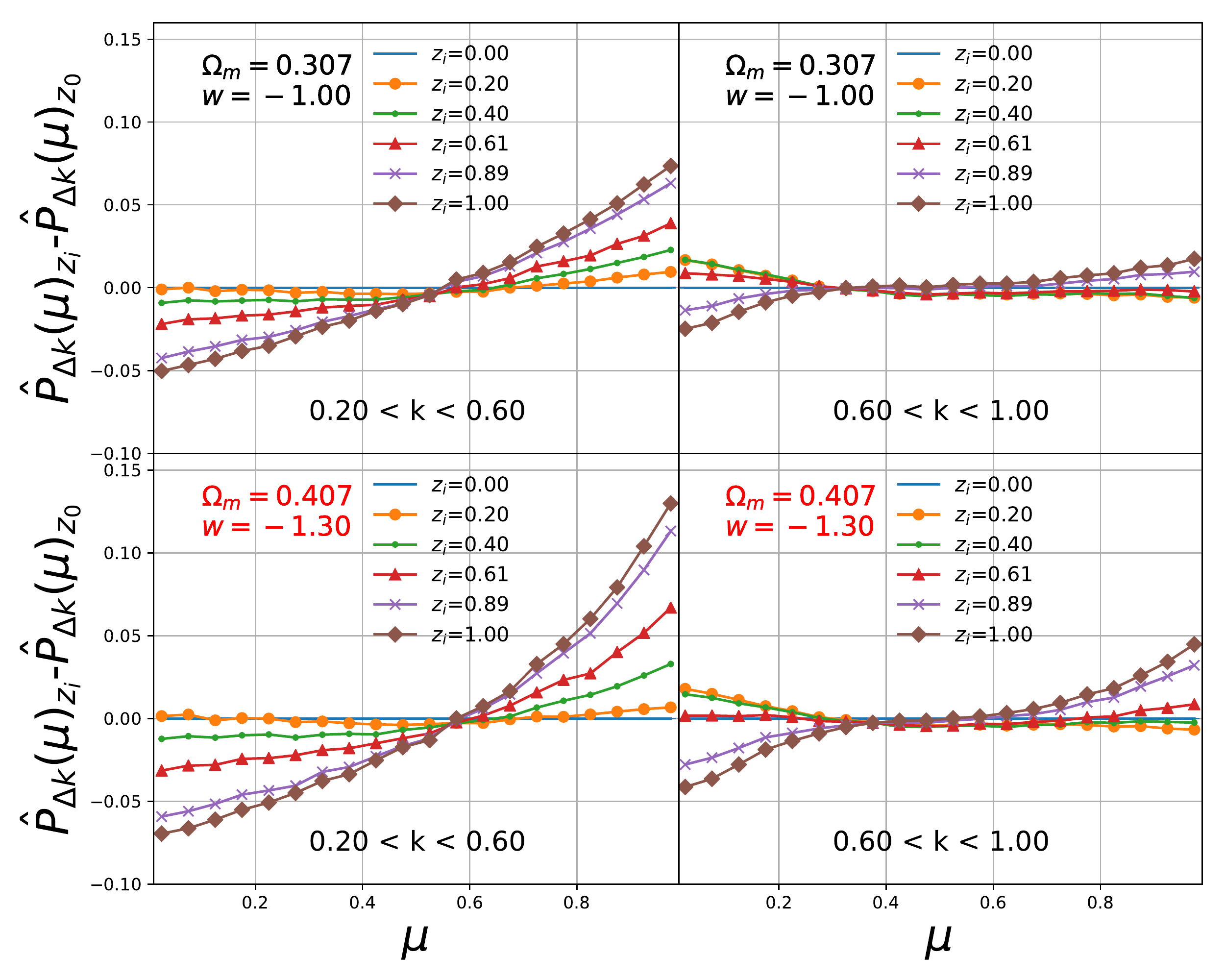}
      \includegraphics[width=8cm]{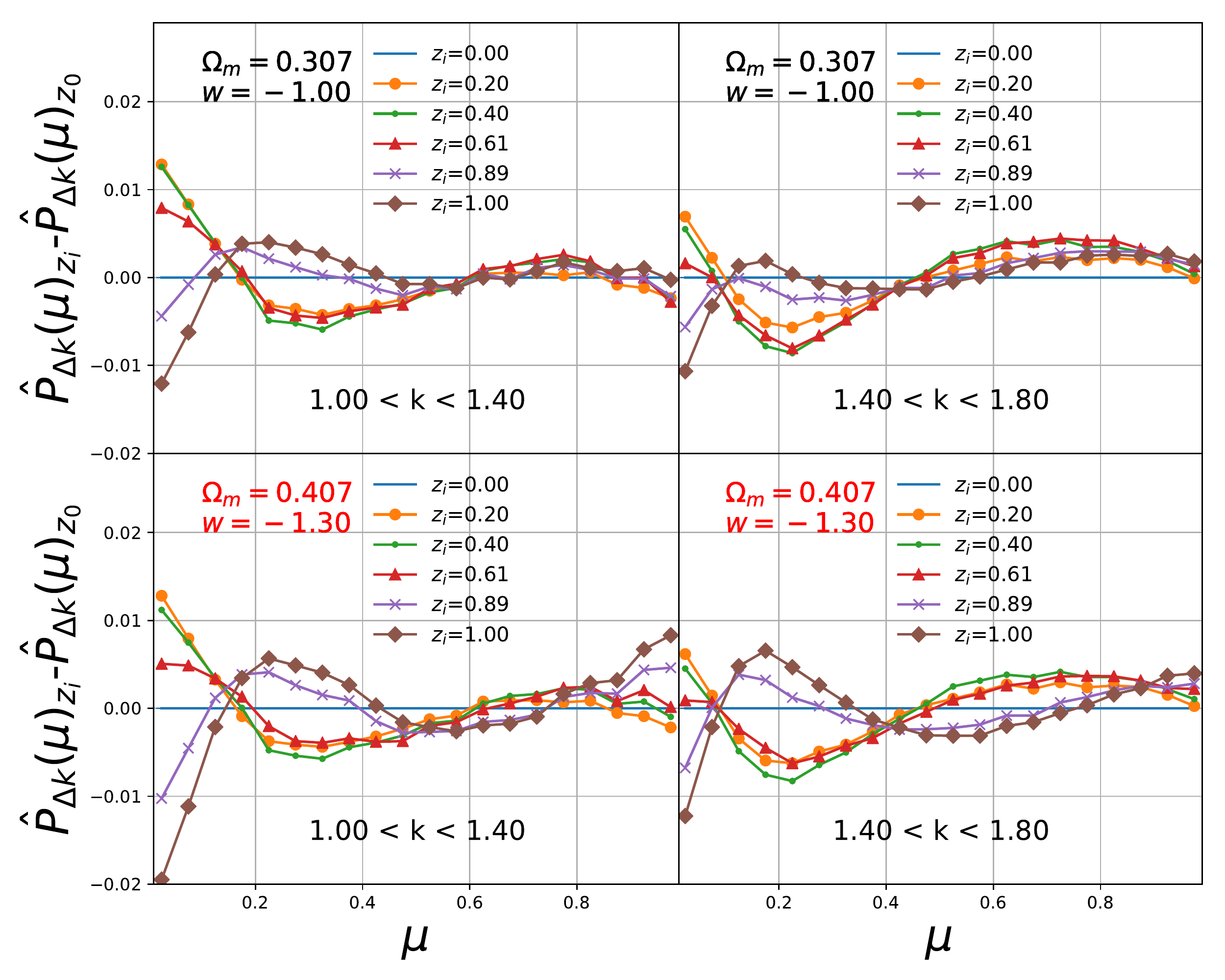}
      \caption{The redshift evolution of $\hat P_{\Delta k}(\mu)$ in cases that
      the range of $k$ chosen as (0.2,0.6), (0.6,1), (1,1.4) and (1.4,1.8) in units of $h\ \rm Mpc^{-1}$, respectively.
      The tomographic AP method works well throughout the regime of $k$.}
      \label{fig:large}
      \end{figure*}

            \begin{figure}[htbp]
      \includegraphics[width=8cm]{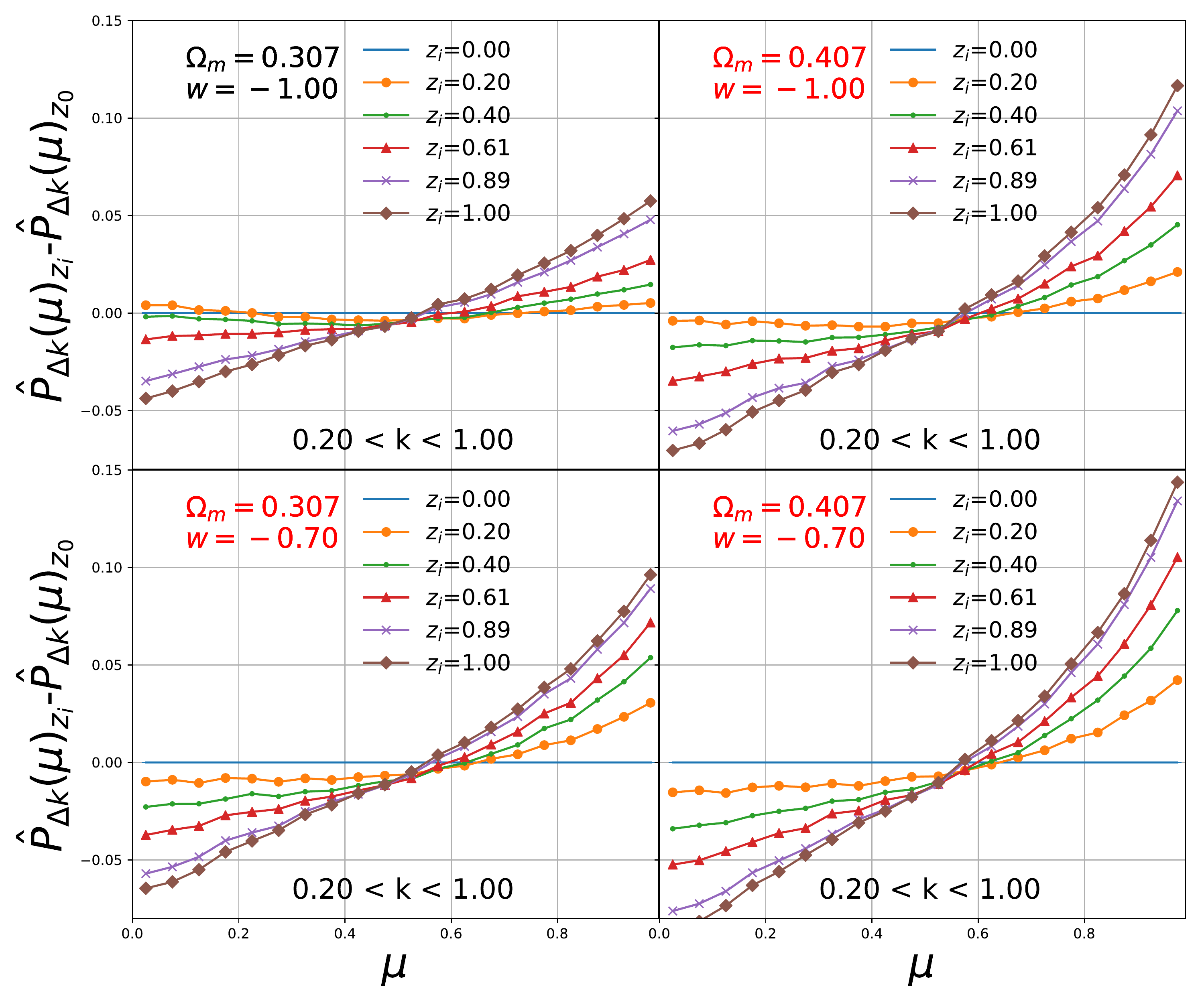}
      \caption{The redshift evolution of $\hat P_{\Delta k}(\mu)$
      measured in the underlying true cosmology and three wrong cosmologies.
      We find significantly larger evolution in all wrong cosmologies.}
      \label{fig:compare}
      \end{figure}

\section{Results}\label{result}

In this work, we compute the 2D PS of the mock data,
using different snapshots and also different assumptions of cosmologies.

\subsection{The 2D PS at different redshifts}

Figure \ref{fig:contour} shows the 2D contour map of the power spectra $P(k,\mu)$
at six redshifts.
The regions $k\in (0.2,1.0)h\ \rm Mpc^{-1}$ are marked by the two dashed lines.

In the $k\lesssim0.2$ regions the pattern of $P(k,\mu)$
behaves as its linear behavior $P\varpropto (1+\beta \mu^2)^2$ \citep{kaiser1987clustering},
where $\beta\equiv b f$ and $b$, $f$ is the bias and the growth rate, respectively.
This relation completely breaks down at $k\gtrsim0.5$,
where we see the PS maximizes at $\mu\sim0$,
reaches its minimal value at $\mu\approx0.6$,
and possesses another peak-value at $\mu\sim1$.

From the plot we find the PS measured at the six redshifts look rather close to each other.
Even though it is difficult to precisely model the
$P(k,\mu)$ in the non-linear region
still we can make use of this redshift ``invariance'' to conduct a cosmological analysis.

\subsection{$\hat{P}_{\Delta k}(\mu)$ in the True Cosmology}

The integrated PS $\hat{P}_{\Delta k}(\mu)$ are plotted in Figure \ref{fig:smallprime},
where we show the results in the fiducial cosmology of the BigMD simulation
as well as a wrong cosmology $\Omega_m=0.407$, $w=-1.30$.
From left to right, we use integration intervals
of (0.2,0.4), (0.4,0.6), (0.6,0.8) and $(0.8,1.0)\ h\ \rm Mpc^{-1}$, respectively.
We find that:

\begin{itemize}
\item
In the first panel ($k\in(0.2,0.4)\ h\ \rm Mpc^{-1}$),
the curves only have a rising trend along $\mu$,
suggesting that their behaviours are dominated by the Kaiser effect \cite{kaiser1987clustering}.
\item
In the second panel ($k\in(0.4,0.6)\ h\ \rm Mpc^{-1}$),
the curves decline at $\mu\lesssim0.55$.
This phenomenon is caused by the FOG effect.
On smaller scales, the FOG effects will dominate 
the linear Kaiser effects over 
almost all $\mu$ values.
However, at large $\mu$ values ($>0.5$) the 
higher-order nonlinear terms 
play an important role and 
can even force the curves to turn over (see \citep{Zheng:2016zxc} for details).
\item
In the last two panels ($k\in(0.6,0.8),\ (0.8,1.0)\ h\ \rm Mpc^{-1}$),
as $k$ becomes larger,
the impact of the FOG effect becomes stronger.
This makes the declining trend in $\mu\lesssim0.6$ stronger,
and the rising trend in $0.6<\mu<1.0$ weaker.
\end{itemize}

In Figure \ref{fig:small} we plotted the redshift evolution of
$\hat P_{\Delta k}(\mu)$, quantified as
\begin{equation}
\delta \hat P_{\Delta k}(\mu,z_1,z_2) \equiv \hat P_{\Delta k}(\mu,z_1) - \hat P_{\Delta k}(\mu,z_2)
\end{equation}
where we choose $z_0=0$ and $z_i =0, 0.2, 0.4, 0.61, 0.89, 1$.
Results of the true Cosmology (upper panel)
show weak and non-zero redshift evolutions in all range of $k$.
The evolutions become smaller as we increase the values of $k$.

\subsection{$\hat{P}_{\Delta k}(\mu)$ and its Redshift Evolution in the Wrong Cosmology}

      \begin{figure*}[htbp]
      \centering

      \includegraphics[width=8cm,height=7cm]{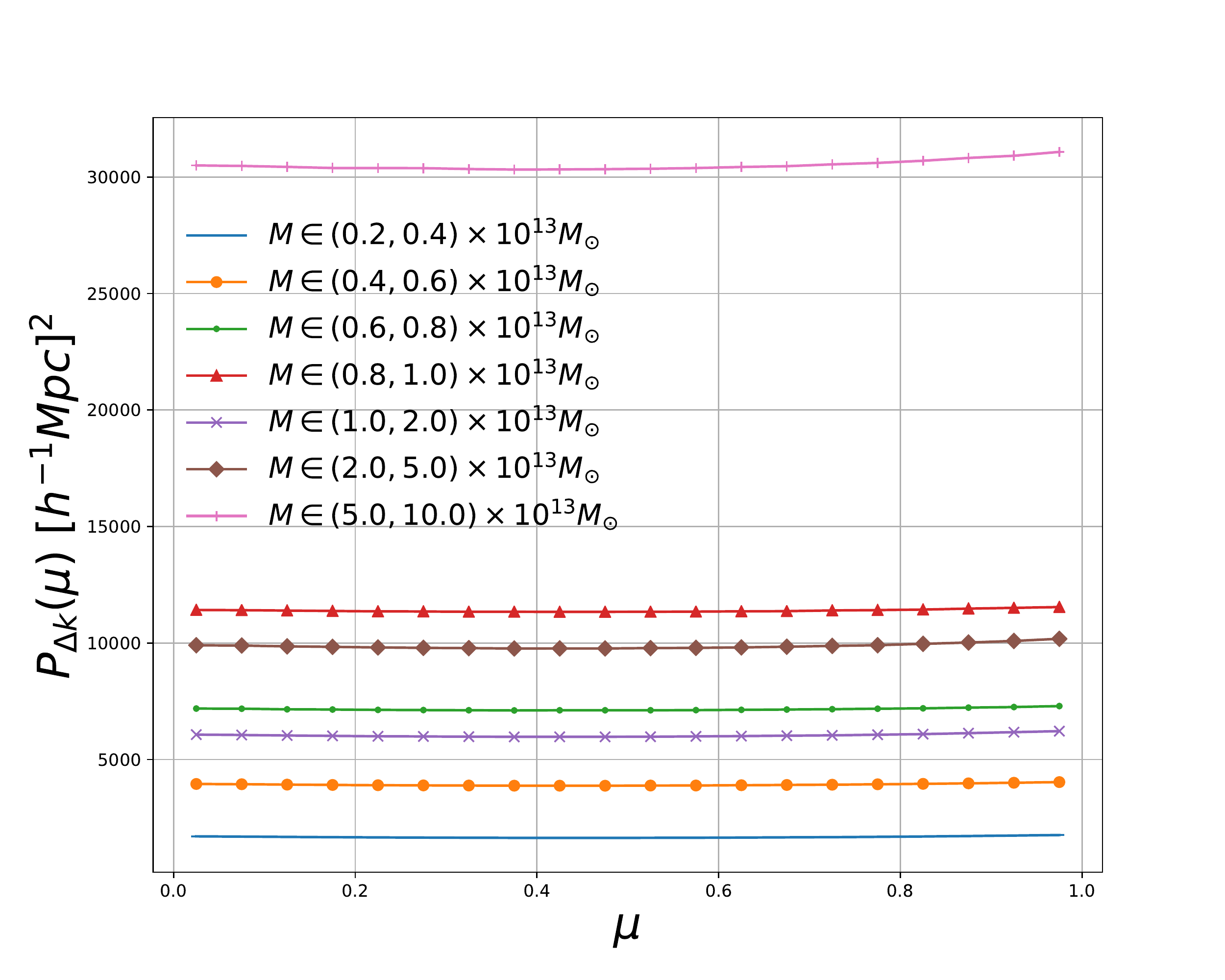}\includegraphics[width=8cm,height=7cm]{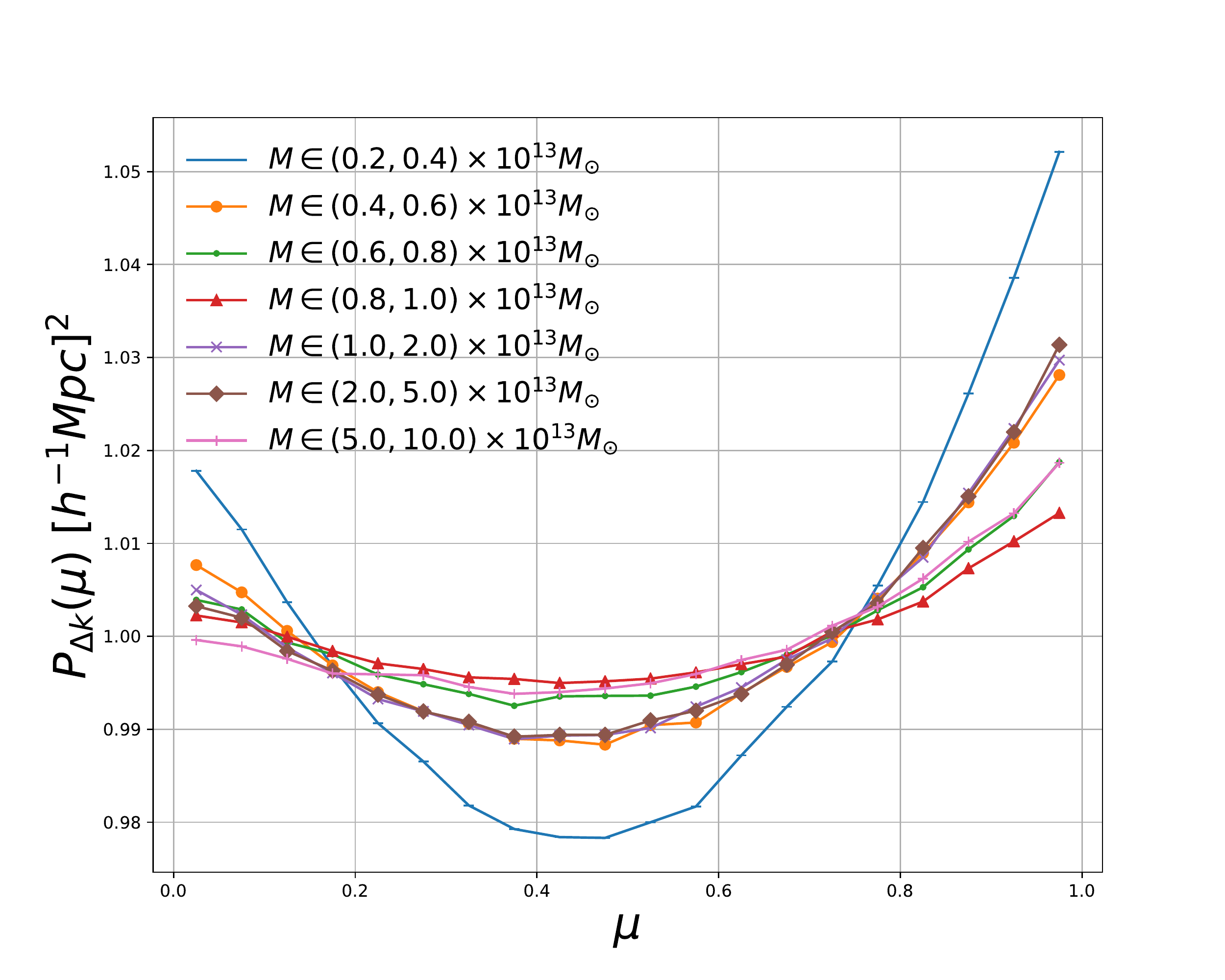}
      \caption{
      The influence of halo bias on the statistical quantity $P_{\Delta k}(\mu)$.
      When changing the mass of halos in range of $2\times 10^{12}-10^{14} M_\odot$,
      we detect significant change in the clustering strength (left panel).
      After the normalization, $\hat P_{\Delta k}(\mu)$ becomes rather insensitive to the halo mass.}
      \label{fig:bias}
      \end{figure*}

In the lower panels Figure\ref{fig:smallprime} and \ref{fig:small}
we plot $\hat{P}_{\Delta k}(\mu)$ and its redshift evolution in a wrong cosmology $\Omega_m=0.407$, $w=-1.30$.
Figure \ref{fig_xy} shows that that,
at $z<0.2$ ($z>0.2$), this cosmology creates compression (stretch) along the LOS,
and the magnitude of compression/stretch largely depends on the redshift.

Correspondingly, in the lower panel of Figure \ref{fig:smallprime},
we see an extra tilt in the PS --
compared with the true cosmology results,
in the wrong cosmology the $\hat{P}_{\Delta k}(\mu)$s have smaller (larger) values in the region of $\mu\lesssim0.5$ ($\mu\gtrsim0.5)$.
The redshift evolution of the anisotropy manifests itself as a redshift evolution of $\hat{P}_{\Delta k}(\mu)$.
In the lower panel Figure \ref{fig:small}, we find that the wrong cosmology
yields larger $\hat{P}_{\Delta k}(\mu,z_i) - \hat{P}_{\Delta k}(\mu,z_0)$ than the true cosmology results
at $z>0.2$.


Using different integration range of $k$,
the additional evolution created by AP is also different.
To have a understanding about their values, we list the value of
$\delta \hat P_{\Delta k}(\mu=1,z_1=0,z_2=1)$ 
in Table \ref{deltaP},
which clearly shows that the wrong cosmology yields to larger values of $\delta \hat P_{\Delta k}$.
Using different choices of $(k_{\rm min},k_{\rm max})$, we find
\begin{equation}
 {\delta \hat P_{\Delta k, \rm wrong\, cosmology}\over\delta \hat P_{\Delta k, \rm true\, cosmology} } \approx1.7-3.6.
\end{equation}



    \begin{table*}
    \caption{\label{deltaP} $\delta \hat P_{\Delta k}(\mu=1,z_1=0,z_2=1)$ in the true cosmology and a wrong cosmology
    $\Omega_m=0.407$, $w=-1.30$
    }\centering
    \begin{tabular*}{\textwidth}{@{}cl*{9}{@{\extracolsep{\fill}}l}}
    \hline\hline
    Integration range of  $k$ ($h\ \rm Mpc^{-1}$)&($0.2,0.4)$&$(0.4,0.6)$&$(0.6,0.8)$&$(0.8,1.0)$&$(0.2,0.6)$&$(0.6,1.0)$ &$(1.4,1.8)$\\
    \hline
    $\delta \hat{P}_{\Delta k, \rm true\, cosmology}$& $8.16\times 10^{-2}$ &$5.73\times 10^{-2}$ &$2.49\times 10^{-2}$ &$7.86\times 10^{-3}$ &$7.35\times 10^{-2}$ &$1.74\times 10^{-2}$  &$1.83\times 10^{-3}$\\
    $\delta \hat{P}_{\Delta k, \rm wrong\, cosmology}$& $1.41\times 10^{-1}$ &$1.07\times 10^{-1}$ &$5.82\times 10^{-2}$ &$2.79\times 10^{-2}$ &$1.30\times 10^{-1}$ &$4.48\times 10^{-2}$ &$3.98\times 10^{-3}$\\
    $\delta \hat{P}_{\Delta k, \rm wrong\, cosmology}/\delta \hat{P}_{\Delta k, \rm true\, cosmology}$&1.73&1.87&2.34&3.55&1.77&2.58&2.17\\
    \hline
    \end{tabular*}
    \end{table*}




\subsection{More Tests on the Integration Range and Cosmologies}

To test the method in wider clustering rage we perform a test using $k$ up to 1.8 $h\ \rm Mpc^{-1}$
Figure \ref{fig:large} shows the results in case that we use the integration range of
$k\in$ (0.2,0.6), (0.6,1) (1,1.4) and (1.4,1.8) $h\ \rm Mpc^{-1}$.
The values of  $\Delta \hat{P}_{\Delta k}(\mu)$ and $\delta \Delta \hat{P}_{\Delta k}(\mu)$ decreases as we increase the value of $k$.
In all cases, the wrong cosmology leads to larger redshift evolution than the true cosmology.
So in optimistic case we may be able to use the non-linear regime of $k=1.8$ $h\ \rm Mpc^{-1}$.

To test it in more cosmologies, in Figure \ref{fig:compare} we further plot  $\delta \hat P_{\Delta k}(\mu)$
in four cosmologies of $(\Omega_m, w)=(0.307,-1),\ (0.407,-1),\ (0.307,-0.7)$
and $(0.407,-0.7)$.
We find larger $\delta \hat P_{\Delta k}$ in all wrong cosmologies.
This suggests that the enlarged redshift evolution of $\hat P_{\Delta k}$
is a universal phenomenon which can be found in a large space of wrong parameters.

\subsection{A Test about Halo Bias}

Since now, we are measuring the PS from samples that have a constant
number density $\bar n = 10^{-3}\ (h^{-1}\rm Mpc)^{-3} $.
To test the effect of {\it selection bias },
it is necessary to test the results from samples having different number density.

In Fig.\ref{fig:bias}, we plot the integrated PS $P_{\Delta k}(\mu)$ in 7 subsamples
of the $z=0$ halos lying in different mass ranges, 
including (0.2,0.4), (0.4,0.6), (0.6,0.8), (0.8,1.0), (1.0,2.0), (2.0,5.0) and (5.0,10.0) in units of $10^{13}M_{\odot}$.
We use   $k\in (0.2,1.8)\ h\ \rm Mpc^{-1}$  in all curves.

Clearly, the clustering strength varies significantly among the subsamples,
resulting in $10\% - 1500\%$ difference among their $P_{\Delta k} (\mu)$s (left panel).
In contrast, after the amplitude normalized,
the $\hat P_{\Delta k}(\mu)$s from the difference subsamples are rather similar.
The largest discrepancy appears in the subsample with a mass range $(0.2,0.4) \times 10^{13} M_{\odot}$,
whose results have a $5\%$ difference from the others.
For the other subsamples,
the difference among their $\hat P_{\Delta k}$ is $\lesssim2\%$.




\section{Conclusions and Discussions}\label{conclu}

In this work, we study the feasibility of conducting the tomographic AP test 
using the PS statistics.
Similar to \cite{LI:2015jra}, we quantify the anisotropic clustering by $\hat P_{\Delta k}(\mu)$,
and use its redshift evolution $\delta\hat P_{\Delta k}$ to probe the AP effect.
To test the method, we use dark matter halos from the BigMD simulation,
and created from it a set of constant number density samples having $\bar n=10^{-3} (h^{-1}\rm Mpc)^{-3}$
at redshifts of $0.2,0.4,0.61,0.89$ and $1.0$.
The ``true'' cosmology (i.e. the simulation cosmology) has cosmological parameters of $\Omega_m=0.3071,w=-1$.

In the incorrect cosmologies of $(\Omega_m, w)=(0.407,-1)$, $(0.307,-0.7)$ and $(0.407,-0.7)$,
we measured larger values of $\delta\hat P_{\Delta k}$ than in the true cosmology.
This means that the AP effect manifests itself as a redshift-dependent anisotropy in the clustering of structures,
and is successfully captured by the PS analysis.
Adjusting the integration interval of $k$ in ranges of 0.2-1.8 $h\ \rm Mpc^{-1}$,
we find that the $\delta\hat P_{\Delta k}$ in the cosmology $\Omega_m=0.407,\ w=-1.3$
was enlarged by a factor of approximately 1.7-3.6.

We emphasize that the clustering scales explored in this work and the previous works \cite{LI14,LI:2015jra,Li:2016wbl}
are quasi- or highly non-linear.
Accurate modeling of $\xi(s,\mu)$ or $P(k,\mu)$ in this region is of great difficulty.
Our method provides a way to extract cosmological information in this region
without having their accurate theoretical predictions.

By focusing on much smaller scales the information explored by the tomographic AP method is
intrinsically different and largely independent from those explored by traditional methods such as BAO
(which mainly use $k\lesssim 0.1 h\ \rm Mpc^{-1}$).
A simple check performed in \citep{Zhang2019} showed that the correlation between the two methods is close to zero.

We explore the non-linear clustering scale up to $k=1.8 h\ \rm Mpc^{-1}$,
and find that the tomographic AP method still work well in that regime.
This is tremendously difficult for most current methods.
On smaller scales, the effect of baryons may become important,
and it is not enough to study it just using pure dark matter simulation.

In  Fourier space we may have advantages of mode-decoupling, easier theoretical prediction,
clearer physical meaning, and sometimes faster computational speed.
But we can not claim that it is a better choice than the configuration space just based on these arguments.
An advantage of using the configuration space is that,
the FOG effect is constrained in the $\lesssim 10 h^{-1}\rm \ Mpc$
(in PS, after a Fourier transform the FOG effect spreads out within a large ($k$,$\mu$) range),
so it is easy to control it.
In any case, it is helpful to have the results derived in both Fourier and configuration spaces,
which are cross-checks of each other.

In this proof-of-concept work we have not predicted the power of the method in constraining $\Omega_m$ and $w$,
which requires calculating the covariance matrix of $\delta \hat P_{\Delta k}(\mu)$ using a large number of realizations.
We leave this for future for when we apply the method to real observational data.





\begin{acknowledgments}


We thank Gong-bo Zhao and Yuting Wang for very helpful suggestions.
XDL acknowledges the supported from NSFC grant (No. 11803094).
CGS acknowledges financial support from the National Research Foundation of Korea
(NRF; \#2017R1D1A1B03034900, \#2017R1A2B2004644 and \#2017R1A4A1015178).

ZL was supported by the Project for New faculty of Shanghai JiaoTong University
(AF0720053),
the National Science Foundation of China (No.  11533006, 11433001)
and the National Basic Research Program of China (973 Program 2015CB857000).

The CosmoSim database used in this paper is a service by the 
Leibniz-Institute for Astrophysics Potsdam (AIP).
The MultiDark database was developed in cooperation with 
the Spanish MultiDark Consolider Project CSD2009-00064.

\end{acknowledgments}

\bibliographystyle{aasjournal}

\begin{thebibliography}{}
\expandafter\ifx\csname natexlab\endcsname\relax\def\natexlab#1{#1}\fi

\bibitem[{{Alam} {et~al.}(2017){Alam}, {Ata}, {Bailey}, {Beutler}, {Bizyaev},
  {Blazek}, {Bolton}, {Brownstein}, {Burden}, {Chuang}, {Comparat}, {Cuesta},
  {Dawson}, {Eisenstein}, {Escoffier}, {Gil-Mar{\'{\i}}n}, {Grieb}, {Hand},
  {Ho}, {Kinemuchi}, {Kirkby}, {Kitaura}, {Malanushenko}, {Malanushenko},
  {Maraston}, {McBride}, {Nichol}, {Olmstead}, {Oravetz}, {Padmanabhan},
  {Palanque-Delabrouille}, {Pan}, {Pellejero-Ibanez}, {Percival}, {Petitjean},
  {Prada}, {Price-Whelan}, {Reid}, {Rodr{\'{\i}}guez-Torres}, {Roe}, {Ross},
  {Ross}, {Rossi}, {Rubi{\~n}o-Mart{\'{\i}}n}, {Saito}, {Salazar-Albornoz},
  {Samushia}, {S{\'a}nchez}, {Satpathy}, {Schlegel}, {Schneider},
  {Sc{\'o}ccola}, {Seo}, {Sheldon}, {Simmons}, {Slosar}, {Strauss}, {Swanson},
  {Thomas}, {Tinker}, {Tojeiro}, {Maga{\~n}a}, {Vazquez}, {Verde}, {Wake},
  {Wang}, {Weinberg}, {White}, {Wood-Vasey}, {Y{\`e}che}, {Zehavi}, {Zhai}, \&
  {Zhao}}]{alam2017clustering}
{Alam}, S., {Ata}, M., {Bailey}, S., {et~al.} 2017, \mnras, 470, 2617

\bibitem[{Alcock \& Paczynski(1979)}]{Alcock:1979mp}
Alcock, C., \& Paczynski, B. 1979, Nature, 281, 358

\bibitem[{{Anderson} {et~al.}(2012){Anderson}, {Aubourg}, {Bailey}, {Bizyaev},
  {Blanton}, {Bolton}, {Brinkmann}, {Brownstein}, {Burden}, {Cuesta}, {da
  Costa}, {Dawson}, {de Putter}, {Eisenstein}, {Gunn}, {Guo}, {Hamilton},
  {Harding}, {Ho}, {Honscheid}, {Kazin}, {Kirkby}, {Kneib}, {Labatie},
  {Loomis}, {Lupton}, {Malanushenko}, {Malanushenko}, {Mandelbaum}, {Manera},
  {Maraston}, {McBride}, {Mehta}, {Mena}, {Montesano}, {Muna}, {Nichol},
  {Nuza}, {Olmstead}, {Oravetz}, {Padmanabhan}, {Palanque-Delabrouille}, {Pan},
  {Parejko}, {P{\^a}ris}, {Percival}, {Petitjean}, {Prada}, {Reid}, {Roe},
  {Ross}, {Ross}, {Samushia}, {S{\'a}nchez}, {Schlegel}, {Schneider},
  {Sc{\'o}ccola}, {Seo}, {Sheldon}, {Simmons}, {Skibba}, {Strauss}, {Swanson},
  {Thomas}, {Tinker}, {Tojeiro}, {Maga{\~n}a}, {Verde}, {Wagner}, {Wake},
  {Weaver}, {Weinberg}, {White}, {Xu}, {Y{\`e}che}, {Zehavi}, \&
  {Zhao}}]{anderson2012clustering}
{Anderson}, L., {Aubourg}, E., {Bailey}, S., {et~al.} 2012, \mnras, 427, 3435

\bibitem[{{Ballinger} {et~al.}(1996){Ballinger}, {Peacock}, \&
  {Heavens}}]{Ballinger1996}
{Ballinger}, W.~E., {Peacock}, J.~A., \& {Heavens}, A.~F. 1996, \mnras, 282,
  877

\bibitem[{{Behroozi} {et~al.}(2013){Behroozi}, {Wechsler}, \& {Wu}}]{2013ApJ}
{Behroozi}, P.~S., {Wechsler}, R.~H., \& {Wu}, H.-Y. 2013, \apj, 762, 109

\bibitem[{{Beutler} {et~al.}(2012){Beutler}, {Blake}, {Colless}, {Jones},
  {Staveley-Smith}, {Poole}, {Campbell}, {Parker}, {Saunders}, \&
  {Watson}}]{beutler20116df}
{Beutler}, F., {Blake}, C., {Colless}, M., {et~al.} 2012, \mnras, 423, 3430

\bibitem[{Bianchi {et~al.}(2015)Bianchi, Gil-Maršªn, Ruggeri, \&
  Percival}]{Bianchi:2015oia}
Bianchi, D., Gil-Maršªn, H., Ruggeri, R., \& Percival, W.~J. 2015, Mon. Not.
  Roy. Astron. Soc., 453, L11

\bibitem[{{Blake} {et~al.}(2011{\natexlab{a}}){Blake}, {Glazebrook}, {Davis},
  {Brough}, {Colless}, {Contreras}, {Couch}, {Croom}, {Drinkwater}, {Forster},
  {Gilbank}, {Gladders}, {Jelliffe}, {Jurek}, {Li}, {Madore}, {Martin},
  {Pimbblet}, {Poole}, {Pracy}, {Sharp}, {Wisnioski}, {Woods}, {Wyder}, \&
  {Yee}}]{blake2011wigglezb}
{Blake}, C., {Glazebrook}, K., {Davis}, T.~M., {et~al.} 2011{\natexlab{a}},
  \mnras, 418, 1725

\bibitem[{{Blake} {et~al.}(2011{\natexlab{b}}){Blake}, {Brough}, {Colless},
  {Contreras}, {Couch}, {Croom}, {Davis}, {Drinkwater}, {Forster}, {Gilbank},
  {Gladders}, {Glazebrook}, {Jelliffe}, {Jurek}, {Li}, {Madore}, {Martin},
  {Pimbblet}, {Poole}, {Pracy}, {Sharp}, {Wisnioski}, {Woods}, {Wyder}, \&
  {Yee}}]{blake2011wigglez}
{Blake}, C., {Brough}, S., {Colless}, M., {et~al.} 2011{\natexlab{b}}, \mnras,
  415, 2876

\bibitem[{{Colless} {et~al.}(2003){Colless}, {Peterson}, {Jackson}, {Peacock},
  {Cole}, {Norberg}, {Baldry}, {Baugh}, {Bland-Hawthorn}, {Bridges}, {Cannon},
  {Collins}, {Couch}, {Cross}, {Dalton}, {De Propris}, {Driver}, {Efstathiou},
  {Ellis}, {Frenk}, {Glazebrook}, {Lahav}, {Lewis}, {Lumsden}, {Maddox},
  {Madgwick}, {Sutherland}, \& {Taylor}}]{2df:Colless:2003wz}
{Colless}, M., {Peterson}, B.~A., {Jackson}, C., {et~al.} 2003, arXiv
  Astrophysics e-prints, astro-ph/0306581

\bibitem[{{Eisenstein} {et~al.}(2005){Eisenstein}, {Zehavi}, {Hogg},
  {Scoccimarro}, {Blanton}, {Nichol}, {Scranton}, {Seo}, {Tegmark}, {Zheng},
  {Anderson}, {Annis}, {Bahcall}, {Brinkmann}, {Burles}, {Castander},
  {Connolly}, {Csabai}, {Doi}, {Fukugita}, {Frieman}, {Glazebrook}, {Gunn},
  {Hendry}, {Hennessy}, {Ivezi{\'c}}, {Kent}, {Knapp}, {Lin}, {Loh}, {Lupton},
  {Margon}, {McKay}, {Meiksin}, {Munn}, {Pope}, {Richmond}, {Schlegel},
  {Schneider}, {Shimasaku}, {Stoughton}, {Strauss}, {SubbaRao}, {Szalay},
  {Szapudi}, {Tucker}, {Yanny}, \& {York}}]{Eisenstein:2005su}
{Eisenstein}, D.~J., {Zehavi}, I., {Hogg}, D.~W., {et~al.} 2005, \apj, 633, 560

\bibitem[{Hand {et~al.}(2018)Hand, Feng, Beutler, Li, Modi, Seljak, \&
  Slepian}]{Hand:2017pqn}
Hand, N., Feng, Y., Beutler, F., {et~al.} 2018, Astron. J., 156, 160

\bibitem[{{Kaiser}(1987)}]{kaiser1987clustering}
{Kaiser}, N. 1987, \mnras, 227, 1

\bibitem[{Kim {et~al.}(2009)Kim, Park, Gott~III, \& Dubinski}]{kim2009horizon}
Kim, J., Park, C., Gott~III, J.~R., \& Dubinski, J. 2009, The Astrophysical
  Journal, 701, 1547

\bibitem[{Kim {et~al.}(2015)Kim, Park, L'Huillier, \& Hong}]{kim2015horizon}
Kim, J., Park, C., L'Huillier, B., \& Hong, S.~E. 2015, arXiv preprint
  arXiv:1508.05107

\bibitem[{{Klypin} {et~al.}(2016){Klypin}, {Yepes}, {Gottl{\"o}ber}, {Prada},
  \& {He{\ss}}}]{2016MNRAS.457.4340K}
{Klypin}, A., {Yepes}, G., {Gottl{\"o}ber}, S., {Prada}, F., \& {He{\ss}}, S.
  2016, \mnras, 457, 4340

\bibitem[{{Lavaux} \& {Wandelt}(2012)}]{lavaux2012precision}
{Lavaux}, G., \& {Wandelt}, B.~D. 2012, \apj, 754, 109

\bibitem[{{Li} {et~al.}(2011){Li}, {Li}, {Wang}, \& {Wang}}]{miao2011dark}
{Li}, M., {Li}, X.-D., {Wang}, S., \& {Wang}, Y. 2011, Communications in
  Theoretical Physics, 56, 525

\bibitem[{{Li} {et~al.}(2019){Li}, {Miao}, {Wang}, {Zhang}, {Fang}, {Luo},
  {Huang}, \& {Li}}]{LI19}
{Li}, X.-D., {Miao}, H., {Wang}, X., {et~al.} 2019, \apj, 875, 92

\bibitem[{Li {et~al.}(2014)Li, Park, Forero-Romero, \& Kim}]{LI14}
Li, X.-D., Park, C., Forero-Romero, J.~E., \& Kim, J. 2014, Astrophys. J., 796,
  137

\bibitem[{Li {et~al.}(2015)Li, Park, Sabiu, \& Kim}]{LI:2015jra}
Li, X.-D., Park, C., Sabiu, C.~G., \& Kim, J. 2015, Mon. Not. Roy. Astron.
  Soc., 450, 807

\bibitem[{Li {et~al.}(2016)Li, Park, Sabiu, Park, Weinberg, Schneider, Kim, \&
  Hong}]{Li:2016wbl}
Li, X.-D., Park, C., Sabiu, C.~G., {et~al.} 2016, Astrophys. J., 832, 103

\bibitem[{Li {et~al.}(2018)Li, Sabiu, Park, Wang, Zhao, Park, Shafieloo, Kim,
  \& Hong}]{LI:2018nlh}
Li, X.-D., Sabiu, C.~G., Park, C., {et~al.} 2018, Astrophys. J., 856, 88

\bibitem[{{Mao} {et~al.}(2017){Mao}, {Berlind}, {Scherrer}, {Neyrinck},
  {Scoccimarro}, {Tinker}, {McBride}, \& {Schneider}}]{Qingqing2016}
{Mao}, Q., {Berlind}, A.~A., {Scherrer}, R.~J., {et~al.} 2017, \apj, 835, 160

\bibitem[{{Matsubara} \& {Suto}(1996)}]{matsubara1996cosmological}
{Matsubara}, T., \& {Suto}, Y. 1996, \apjl, 470, L1

\bibitem[{{Outram} {et~al.}(2004){Outram}, {Shanks}, {Boyle}, {Croom}, {Hoyle},
  {Loaring}, {Miller}, \& {Smith}}]{outram20042df}
{Outram}, P.~J., {Shanks}, T., {Boyle}, B.~J., {et~al.} 2004, \mnras, 348, 745

\bibitem[{Park {et~al.}(2019)Park, Park, Sabiu, Li, Hong, Kim, Tonegawa, \&
  Zheng}]{Park:2019mvn}
Park, H., Park, C., Sabiu, C.~G., {et~al.} 2019, arXiv:1904.05503

\bibitem[{{Percival} {et~al.}(2007){Percival}, {Cole}, {Eisenstein}, {Nichol},
  {Peacock}, {Pope}, \& {Szalay}}]{Percival:2007yw}
{Percival}, W.~J., {Cole}, S., {Eisenstein}, D.~J., {et~al.} 2007, \mnras, 381,
  1053

\bibitem[{{Perlmutter} {et~al.}(1999){Perlmutter}, {Aldering}, {Goldhaber},
  {Knop}, {Nugent}, {Castro}, {Deustua}, {Fabbro}, {Goobar}, {Groom}, {Hook},
  {Kim}, {Kim}, {Lee}, {Nunes}, {Pain}, {Pennypacker}, {Quimby}, {Lidman},
  {Ellis}, {Irwin}, {McMahon}, {Ruiz-Lapuente}, {Walton}, {Schaefer}, {Boyle},
  {Filippenko}, {Matheson}, {Fruchter}, {Panagia}, {Newberg}, {Couch}, \&
  {Project}}]{perlmutter1999measurements}
{Perlmutter}, S., {Aldering}, G., {Goldhaber}, G., {et~al.} 1999, \apj, 517,
  565

\bibitem[{{Ramanah} {et~al.}(2019){Ramanah}, {Lavaux}, {Jasche}, \&
  {Wandelt}}]{KR2018}
{Ramanah}, D.~K., {Lavaux}, G., {Jasche}, J., \& {Wandelt}, B.~D. 2019, \aap,
  621, A69

\bibitem[{{Riess} {et~al.}(1998){Riess}, {Filippenko}, {Challis},
  {Clocchiatti}, {Diercks}, {Garnavich}, {Gilliland}, {Hogan}, {Jha},
  {Kirshner}, {Leibundgut}, {Phillips}, {Reiss}, {Schmidt}, {Schommer},
  {Smith}, {Spyromilio}, {Stubbs}, {Suntzeff}, \&
  {Tonry}}]{riess1998observational}
{Riess}, A.~G., {Filippenko}, A.~V., {Challis}, P., {et~al.} 1998, \aj, 116,
  1009

\bibitem[{{Ryden}(1995)}]{ryden1995measuring}
{Ryden}, B.~S. 1995, \apj, 452, 25

\bibitem[{Scoccimarro(2015)}]{Scoccimarro:2015bla}
Scoccimarro, R. 2015, Phys. Rev., D92, 083532

\bibitem[{Wang(2008)}]{Wang:2008zh}
Wang, Y. 2008, Phys. Rev., D77, 123525

\bibitem[{{Weinberg} {et~al.}(2013){Weinberg}, {Mortonson}, {Eisenstein},
  {Hirata}, {Riess}, \& {Rozo}}]{weinberg2013observational}
{Weinberg}, D.~H., {Mortonson}, M.~J., {Eisenstein}, D.~J., {et~al.} 2013,
  \physrep, 530, 87

\bibitem[{{Weinberg}(1989)}]{weinberg1989cosmological}
{Weinberg}, S. 1989, Reviews of Modern Physics, 61, 1

\bibitem[{{York} {et~al.}(2000){York}, {Adelman}, {Anderson}, {Anderson},
  {Annis}, {Bahcall}, {Bakken}, {Barkhouser}, {Bastian}, {Berman}, {Boroski},
  {Bracker}, {Briegel}, {Briggs}, {Brinkmann}, {Brunner}, {Burles}, {Carey},
  {Carr}, {Castander}, {Chen}, {Colestock}, {Connolly}, {Crocker}, {Csabai},
  {Czarapata}, {Davis}, {Doi}, {Dombeck}, {Eisenstein}, {Ellman}, {Elms},
  {Evans}, {Fan}, {Federwitz}, {Fiscelli}, {Friedman}, {Frieman}, {Fukugita},
  {Gillespie}, {Gunn}, {Gurbani}, {de Haas}, {Haldeman}, {Harris}, {Hayes},
  {Heckman}, {Hennessy}, {Hindsley}, {Holm}, {Holmgren}, {Huang}, {Hull},
  {Husby}, {Ichikawa}, {Ichikawa}, {Ivezi{\'c}}, {Kent}, {Kim}, {Kinney},
  {Klaene}, {Kleinman}, {Kleinman}, {Knapp}, {Korienek}, {Kron}, {Kunszt},
  {Lamb}, {Lee}, {Leger}, {Limmongkol}, {Lindenmeyer}, {Long}, {Loomis},
  {Loveday}, {Lucinio}, {Lupton}, {MacKinnon}, {Mannery}, {Mantsch}, {Margon},
  {McGehee}, {McKay}, {Meiksin}, {Merelli}, {Monet}, {Munn}, {Narayanan},
  {Nash}, {Neilsen}, {Neswold}, {Newberg}, {Nichol}, {Nicinski}, {Nonino},
  {Okada}, {Okamura}, {Ostriker}, {Owen}, {Pauls}, {Peoples}, {Peterson},
  {Petravick}, {Pier}, {Pope}, {Pordes}, {Prosapio}, {Rechenmacher}, {Quinn},
  {Richards}, {Richmond}, {Rivetta}, {Rockosi}, {Ruthmansdorfer}, {Sandford},
  {Schlegel}, {Schneider}, {Sekiguchi}, {Sergey}, {Shimasaku}, {Siegmund},
  {Smee}, {Smith}, {Snedden}, {Stone}, {Stoughton}, {Strauss}, {Stubbs},
  {SubbaRao}, {Szalay}, {Szapudi}, {Szokoly}, {Thakar}, {Tremonti}, {Tucker},
  {Uomoto}, {Vanden Berk}, {Vogeley}, {Waddell}, {Wang}, {Watanabe},
  {Weinberg}, {Yanny}, {Yasuda}, \& {SDSS Collaboration}}]{york2000sloan}
{York}, D.~G., {Adelman}, J., {Anderson}, Jr., J.~E., {et~al.} 2000, \aj, 120,
  1579

\bibitem[{Zhang {et~al.}(2019)Zhang, Gu, Wang, Li, Sabiu, Park, Miao, Luo,
  Fang, \& Li}]{Zhang2019}
Zhang, Z., Gu, G., Wang, X., {et~al.} 2019, Astrophys. J., 878, 137

\bibitem[{Zheng \& Song(2016)}]{Zheng:2016zxc}
Zheng, Y., \& Song, Y.-S. 2016, JCAP, 1608, 050

\end{thebibliography}

\end{document}